\documentstyle[prl,aps,multicol,epsf]{revtex}
\tighten
\begin{document}
\draft

\title{Interaction Constants and Dynamic Conductance of a Gated Wire}
\author{Ya. M. Blanter$^{(1)}$, F.W.J. Hekking$^{(2)}$, and M. 
B\"uttiker$^{(1)}$}
\address{$^{(1)}$D\'epartement de Physique Th\'eorique, Universit\'e 
de Gen\`eve, CH-1211, Gen\`eve 4, Switzerland\\
$^{(2)}$Cavendish Laboratory, University of Cambridge, Madingley Road, 
Cambridge CB3 0HE, United Kingdom\\}
\date{\today}
\maketitle

\begin{abstract}
We show that the interaction constant governing 
the long-range electron-electron interaction in a quantum wire
coupled to two reservoirs and capacitively coupled to a gate 
can be determined by a low frequency measurement. 
We present a self-consistent, charge and current 
conserving theory of the full conductance matrix. The collective 
excitation spectrum consists of plasma modes with a relaxation
rate which increases with the interaction strength and is 
inversely proportional to the length of the wire. The interaction
parameter is determined by the first two coefficients of the 
out-of-phase component of the dynamic conductance measured at the gate. 
\end{abstract}


\begin{multicols}{2}
The comparison of interacting electron theories
and experiments often suffers from the fact that the interaction
constants are not known. In particular, this is true for interacting
quantum wires, where Luttinger models with a wide range of coupling
parameters are discussed~\cite{Haldane81Kane92,Tarucha95Yacoby96}.
Moreover, a single experiment is often not 
sufficient to determine the coupling constant. 
Thus,
as will be shown, the capacitance (per unit length) of a wire above
a back gate is related to the interaction parameter $g$ via
\begin{equation} \label{cmu}
c_\mu = g^{2} {e^{2}} \nu_F . 
\end{equation}
Since the density of states $\nu_F = 2/h v_F$
(evaluated at constant interaction potential~\cite{Buettiker93})
is generally not known, a capacitance measurement alone cannot
determine the interaction constant. Here we propose to investigate the
frequency dependence of current induced into the gate. Compared to the
measurement of a frequency dependent conductance at a direct contact, 
the measurement of the gate current can be performed at relatively 
small frequencies in the kHz range, since the frequency response is
not on top of a possibly large dc-conductance. Recently considerable
advances have been made in the high precision frequency measurement of
the complex ac conductance~\cite{Pieper94} and the frequency 
dependent noise~\cite{Schoelkopf97} of 
mesoscopic conductors. In this Letter we consider a simple model 
system -- a perfect ballistic wire coupled capacitively to a gate
and connected to two electron reservoirs -- and calculate the 
dynamic conductance matrix. While the dc-conductance in this system is
quantized and thus provides no information on the interaction, the
dynamic conductance is a sensitive function of the interaction
strength.   
 
A conceptually important point which needs to be addressed 
in solving this problem is the coupling of an interacting 
wire to electron reservoirs. 
Previous works~\cite{Maslov95Ponomarenko95Safi95} 
have proposed a purely one-dimensional model in which the interaction
changes from a value $g < 1$ (characteristic of interactions)
to a value $g=1$ (characterizing a system without interactions) at the
transition from the ballistic wire to the electron reservoir. 
Another more recent proposal~\cite{Egger96} consists of a radiative
boundary condition in which the electron density is proportional to
the applied voltage. In this work we use a concept of reservoirs based
on the {\it electrochemical} nature of electric
transport\cite{Buettiker93}. The electron density in the wire is the
sum of two terms: a chemical density, which follows the chemical 
potential of the reservoir from which the 
carriers are injected into the wire, and an induced density, 
which results from the (long range) Coulomb screening of the 
injected charge. Indeed, from a screening point of view 
electrons in a reservoir are not free: an increase of the 
electrochemical potential is followed by an equal increase in 
the electrostatic potential. That corresponds to strong interaction
(very effective three-dimensional screening) in a reservoir rather
than to a non-interacting one-dimensional (1D) model with $g=1$. 
Similarly, the radiative boundary condition employed in
Ref.~\cite{Egger96} is not correct: in a reservoir an increase of a
voltage leaves the local density invariant. These divergent views
arise from the fact that interactions play a role on very different
length scales~\cite{Nozieres}. Different interaction parameters 
must be used to describe long range and short range effects~\cite{Wen}. 
Conceptually, Ref.~\cite{Alekseev97}, 
which describes the reservoirs by the
charges, conjugate to the chemical reservoir potentials, 
is closest to our approach. 

Ballistic single mode wires~\cite{Haldane81Kane92,Tarucha95Yacoby96} coupled
to reservoirs are the simplest model system in which these questions
are significant.
The dynamic response of 1D interacting systems has been investigated
previously in the framework of the Luttinger model by several
authors~\cite{Cuniberti96,Ponomarenko96,Sablikov97}. 
The results obtained in these works are, however, 
not charge and current conserving (gauge invariant)~\cite{self-cons}. 
Below we present results for the ac conductance of an interacting
quantum wire, connected to two reservoirs
and capacitively coupled to a gate.
On a length scale large compared to the distance between the wire and 
the gate the interactions can be treated as short ranged. 
Our discussion explicitly includes the effect of the gate. Of interest
is the displacement current which flows from the wire to the gate and
which we propose to measure to determine the interaction constant. 

\begin{figure}
\narrowtext
{\epsfxsize=7cm\epsfysize=3.0cm\centerline{\epsfbox{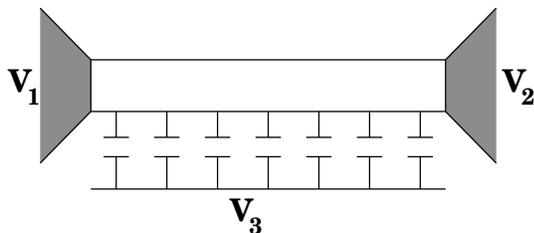}}}
\caption{The 1D wire, connected to two reservoirs and coupled 
capacitively to a gate.}
\label{fig1}
\end{figure}

Consider the system depicted in Fig.~\ref{fig1}, consisting of 
a 1D quantum wire of length $L$, connected to two reservoirs
at $x=0$ and $x=L$. The potential in the left (1) reservoir is 
modulated in time, $V_{1}(t) = V_{1,\omega} e^{-i \omega t}$, whereas 
the potential in the right (2) reservoir is kept constant.   
We treat the interactions in random phase approximation (RPA).
For electron densities which are not too small ($e^{2}/h v_F <
1$, where $v_F$ is the Fermi velocity), RPA is essentially exact. We
note also that even within the strict limits of RPA validity the
interaction constant $g$ may still be made small by appropriate choice
of capacitance between the wire and the gate, see Eq. (\ref{g})
below. 

{\bf Self-consistent potential}. In the absence of interactions, a
potential modulation in the left reservoir injects a bare charge
density  
$\rho_{0,\omega}(x)$ into the wire,
\begin{equation} \label{rho_0}
\rho_{0,\omega}(x) = 
\frac{\nu _{F} V_{1,\omega}}{2} e^{iq_{F}x} ,
\end{equation}
where $\nu _{F} = 2/ h v_F$ is the density of states at the
Fermi level ($v_F$ is the Fermi velocity), and $q_F = \omega /v_F$. 

In the presence of an external potential $\phi _{\omega}(x)/e$, the 
true charge density is
\begin{equation} \label{rho}
\rho_{\omega}(x) 
= 
\rho_{\omega,0}(x) - \int \limits _{0} ^{L} dx'
                     \Pi _{\omega} (x,x') \phi_{\omega}(x'),
\end{equation}
where the polarization kernel is given by (see {\em
e.g.}~\cite{Blanter97}) 
\begin{equation} \label{Pi}
\Pi _{\omega} (x,x') 
=
\nu _{F} \delta(x-x') + \frac{iq_{F} \nu_{F}}{2} e^{iq_{F}|x-x'|}.
\end{equation}
Eq. (\ref{rho}) gives the charge density
as a sum of two contributions:
a chemical one, proportional to the potential $V_{1,\omega}$ of the 
reservoir, and an induced component, proportional to the electrostatic
potential.  

We now take electron-electron interactions into account by determining
the actual potential $\phi_{\omega}(x)$ in the wire
self-consistently, {\em i.e.}, by relating the total charge
$\rho_{\omega}$ on the left hand side of Eq.~(\ref{rho}) to the
potential $\phi_{\omega}$. In order to do this, we need to specify
the electron-electron interaction. In the case of bare
Coulomb interactions, the required relation is the Poisson
equation $\Delta \phi_\omega = -4\pi e^2 \rho_\omega$. For 
short-ranged interactions this relation is different and is 
found as follows: Generally, the interaction 
potential $V(\bbox{r}, \bbox{r'})$ is the Green's
function for the operator equation $\hat{K} V(\bbox{r},\bbox{r'}) =
4\pi e^2 \delta(\bbox{r}-\bbox{r'})$. The same operator $\hat{K}$ also
connects charge $\rho$ and potential $\phi$ via $\hat{K}
\phi(\bbox{r}) = 4\pi e^2 \rho(\bbox{r}).$ 
For bare Coulomb interactions, $\hat{K}$ is the Laplace
operator $\Delta$, and the Poisson equation follows. Here we consider
short range interactions characterized by the interaction strength
$V_0$, $V(x-x') = V_{0} \delta(x-x')$. In this case the operator
$\hat{K}$ is just a multiplication with a constant factor $4\pi
e^2/V_0$. Thus, the potential and charge are connected via 
\begin{equation} \label{modpois}
\phi_{\omega}(x) =
V_{0}\rho_{\omega}(x) ,
\end{equation}
instead of the Poisson equation. At this point it therefore is
quite natural to introduce the capacitance
of the wire per unit length, $c = e^{2}/V_{0}$. Physically, this
corresponds to a single mode quantum wire, formed by
depletion induced by a back-gate with a capacitance $c$ per unit
length, parallel to the wire (see
Fig.~\ref{fig1}). The well-known interaction parameter $g$ of the LL
is then related to the capacitance $c$~\cite{capacitance}
via  
\begin{equation} 
g^2
=
\frac{1}{1+e^{2}\nu_F/c} .
\label{g}
\end{equation}
In particular, the case of the locally charge neutral
wire\cite{Blanter97}, corresponds to $c = 0$ or infinitely strong
point-like interactions ($g=0$). Indeed, the single-channel results of
Ref. \cite{Blanter97} are obtained as the $g \to 0$ limit of the
formulas derived below.  

Now we generalize our approach to the case when the
back-gate is modulated by a potential $V_{3}(t) = V_{3,\omega} e^{-i
\omega t}$ as well. Then the total density of the wire 
contains in addition to the density injected from reservoir $1$
an induced density due to the modulation of the gate.  
The self-consistent potential distribution
$\phi_{\omega}(x)$ along the wire must now be 
found from the equation 
\end{multicols}
\widetext
\vspace*{-0.2truein} \noindent \hrulefill \hspace*{3.6truein}
\begin{equation} \label{eqphi}
\frac{c}{e^2 \nu_F} \left( \phi_{\omega}(x) - V_{3,\omega} \right) =
\frac{V_{1,\omega}}{2} e^{iq_{F}x} 
-\phi_{\omega}(x)
-
\frac{iq_{F}}{2}\int \limits_{0}^{L} dx'
                 e^{iq_{F}|x-x'|}\phi_{\omega}(x') , 
\end{equation}

\begin{multicols}{2} 

\noindent
which is obtained by substituting $\phi_{\omega} (x) - V_{3,\omega}$
in the lhs of Eq.~(\ref{modpois}) and using Eq.~(\ref{rho}).
The solution to Eq.~(\ref{eqphi}) is 
\begin{equation} \label{phi}
\phi_{\omega}(x)
=
V_{3,\omega} +
A^+_{\omega} e^{iqx} + A^-_{\omega} e^{-iqx},
\end{equation}
where $q = g q_{F}$, and
\begin{eqnarray} \label{A}
A^{\pm}_{\omega}
& = &
\pm \frac{(1 \pm g)(1-g^{2})e^{\mp iqL}}{(1+g)^{2}e^{-iqL} -
(1-g)^{2}e^{iqL}} 
\nonumber \\
& \times & \left[ V_{1,\omega}-V_{3,\omega} \left(1-\frac{1 \mp g}{1
\pm g}
e^{\pm iqL} \right) \right].
\end{eqnarray}

{\bf Conductance matrix}. With the true potential we are now in a
position to find the full conductance matrix for the capacitively
coupled wire. The conductance matrix ${\cal G} _{\alpha 
\beta}(\omega)$ relates the current $I_{\alpha,\omega}$ at
contact $\alpha$ to the voltage $V_{\beta,\omega}$ applied at contact 
$\beta$ ($\alpha, \beta = 1,2,3$):
$I_{\alpha,\omega } = {\cal G}_{\alpha \beta}(\omega) 
V_{\beta, \omega}$. 
With the definitions
\begin{equation}
G_{\omega} = 
2g \frac{e^{2}}{h} 
\frac{(1+g)e^{-iqL}+(1-g)e^{iqL}}{(1+g)^{2}e^{-iqL} - (1-g)^{2}e^{iqL}}
\end{equation}
and 
\begin{equation}
\bar{G}_\omega =-\frac{e^{2}}{h} 
\frac{4g}{(1+g)^{2}e^{-iqL} - (1-g)^{2}e^{iqL}},
\end{equation}
the conductance matrix takes a form
\begin{equation}
{\cal G}(\omega)
=
\left(
\begin{array}{ccc}
G_{\omega}  &\bar{G}_\omega &-G_{\omega}  -\bar{G}_\omega \\
\bar{G}_\omega &G_{\omega} &-G_{\omega}  -\bar{G}_\omega \\
-G_{\omega}  -\bar{G}_\omega &-G_{\omega}  -\bar{G}_\omega 
&2G_{\omega}  +2\bar{G}_\omega
\end{array}
\right) .
\label{condmat}
\end{equation}
The matrix ${\cal G}$ has the following properties
\cite{Buettiker93}: First, it is symmetric, which reflects the fact
that the geometry 
considered here is symmetric under the exchange of the left and right
reservoir, and no magnetic field is present. Then, 
$\sum _{\alpha} {\cal G}_{\alpha \beta} = 0$, which re-states current
conservation. Finally, the property $\sum _{\beta} {\cal G}_{\alpha
\beta} = 0$ manifests the fact that a simultaneous 
shift of all potentials $V_{\beta}$ by the same amount does 
not produce any current (gauge invariance). Furthermore, dissipation 
of power requires that the matrix ${\rm Re} \ \cal{G}$ is positive
definite.  
    
In the static limit $\omega=0$ one reproduces the known
result~\cite{Maslov95Ponomarenko95Safi95}: $G = -\bar G = e^2/ h$,
with no current flowing through the gate. Another limiting case is
$g=0$ \cite{Blanter97}, where one finds  $G = -\bar G = (e^2/ h)(1 - i
q_F L/2)^{-1}$. Generally, all the components of the conductance
matrix are oscillating functions of frequency (for $g \ne 0$) with the
period  $2\pi v_F (gL)^{-1}$. In particular, the real part of the
conductance  reaches zero with a period $2\pi v_F (gL)^{-1}$. It has
been suggested that measurement of this period should be used to
determine the interaction
constant~\cite{Ponomarenko96,Sablikov97}. However, this period is a
consequence of the linearization of the spectrum near the Fermi energy
and not really a signature of an interacting system. Furthermore, this
frequency is already in the absence of interactions of the order of an
electron transit frequency and therefore rather high. A much better
strategy consists in analyzing one of the purely capacitive
conductances between the wire and the gate. In particular, we consider
${\cal G}_{33} (\omega) = 2(G_{\omega} + \bar
G_{\omega})$, which we call the gate conductance. The real and
imaginary parts of the frequency dependence of the gate conductance
${\cal G}_{33} (\omega)$ are displayed in Fig.~2. The real part shows
peaks around $\omega = (\pi v_F/g L)(2n+1)$, $n \in \cal{Z}$.
The height of each peak is equal to four times the conductance quantum
$e^2/ h$ (and thus independent of $g$), while the width
decreases with decreasing $g$. In contrast, the imaginary part of
${\cal G}_{33} (\omega)$ changes sign at these points, and exhibits
extrema (sharp ones for small $g$) at the points  
$$\Omega_{n}= \frac{v_F}{gL} \left[
\pi(2n+1) \pm \arccos \frac{1-g^2}{1+g^2} \right].$$
The height of each peak is $2e^2/ h $. 

\begin{figure}
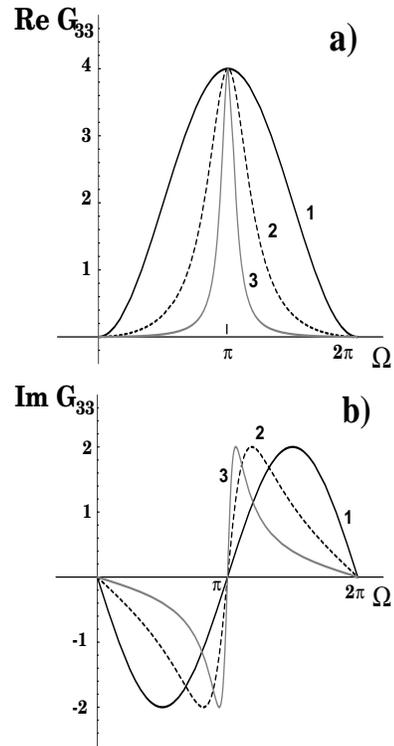

\narrowtext
{\epsfxsize=5.0cm\epsfysize=5.0cm\centerline{\epsfbox{lutt2a.eps}}}
{\epsfxsize=5.0cm\epsfysize=5.0cm\centerline{\epsfbox{lutt2b.eps}}}
\caption{One period of the frequency dependence of the real (a) and
imaginary (b) parts of the gate conductance ${\cal G}_{33} (\omega)$
(in units $e^2/h$). The parameter $g$ is equal to 1 (curve 1), 0.3
(2), and 0.1 (3); the argument is $\Omega = \omega L g/v_F$.}
\label{fig2}
\end{figure}

All elements of the conductance 
matrix ${\cal G}_{\alpha \beta} $ are characterized by
the common denominator $(1+g)^{2}e^{-iqL} - (1-g)^{2}e^{iqL}$, which
has zeros at frequencies 
\begin{equation} \label{poles}
\omega _{n}
=
\frac{v_{F}}{gL}
\left[
n \pi -i \ln \frac{1+g}{1-g} 
\right] .
\end{equation}
Eq. (\ref{poles}) defines the (collective) excitation spectrum of the
quantum wire. For $g=0$ (local charge neutrality) only the  $n=0$ mode
survives, and all other frequencies are pushed up to infinity. This
mode is purely imaginary, and does not correspond to any type of
quasiparticles \cite{Blanter97}. On the other hand, for $g=1$
(non-interacting system) all modes are infinitely damped: Thus, charge
relaxation can only be caused by electron-electron interactions. We
mention that the same modes are obtained in Ref. \cite{Ponomarenko96};
the modes with even $n$ are also obtained in
Ref. \cite{Sablikov97}. Any treatment, whether self-consistent or not,
which at some stage invokes the effective interaction, will exhibit
this frequency spectrum.  

We have now characterized the dynamic conductance and its properties 
over a wide frequency range. But it is the low frequency regime that 
is experimentally most easily accessed. A low frequency measurement
works only if we consider the gate current since at small frequencies 
the ac component of the conductance ${\cal G}_{11}$ represents only a
small deviation from the quantized dc-conductance and is hard
to identify\cite{Chen}. The gate conductance has the following 
low frequency expansion
$${\cal G}_{33} = -i C_{\mu} \omega  
+  R_q C^{2}_{\mu} \omega^{2}  
- i \frac{1-3g^{2}}{3g^{2}} R_q^2 C^{3}_{\mu}
\omega^{3} 
+ \dots$$  
Here $C_\mu = c_\mu L$ is the total electrochemical capacitance
of the wire vis-a-vis the gate, $c_\mu$ is given by Eq. (\ref{cmu}). 
The second order term is determined by the charge relaxation
resistance\cite{BTP93} $R_q = h /4 e^2$ which is
the parallel resistance of two Sharvin-Imry contact resistances of
{\it half} a resistance quantum per contact. It is independent of the
interaction constant.  
The third order term is proportional to the third power 
of the electrochemical capacitance $C_\mu$, but most importantly
it is proportional to a factor
$1/3g^2 - 1$, which is a sensitive function of the interaction
strength. Thus, a measurement which determines 
the out-of-phase (non-dissipative) part of the gate conductance up to
the  the third order in frequency is sufficient to determine the
interaction parameter $g$.

{\bf Conclusions}. We have investigated the ac response of a quantum
wire with short-range interactions. We formulated a self-consistent, 
charge and current conserving approach using RPA. For a wire 
with a density which is not too small, RPA captures the essential 
physics. The boundary condition which couples the density of the
wire to the electron reservoirs is of eletrochemical nature. Due 
to the coupling with the reservoir all the collective modes of 
the system acquire a damping constant. In the present paper the 
one-channel case is considered only. The case
of two channels with the same velocity $v_F$ (corresponding to one
spin-degenerate channel) can be obtained from the above
results simply by replacing the density of states of the one-channel
problem $\nu$ by that appropriate for the spin-degenerate channel
$2\nu$: Spin--charge separation can not be probed by the ac response. 
  
We find that the measurement of the low-frequency, non-dissipative
component of the gate conductance including only its first two leading
coefficients is sufficient to determine the interaction strength.   
Such measurements are very desirable and will provide a strong 
stimulation for further research on the role of electron-electron 
interactions. 

We acknowledge the financial support of the Swiss National Science
Foundation and of the European Community (Contract ERB-CHBI-CT941764).

\end{multicols}

\begin{references}

\bibitem{Haldane81Kane92} F.~D.~M.~Haldane, J. Phys. C {\bf 14}, 2585
(1981); C.~L.~Kane and M.~P.~A.~Fisher, Phys. Rev. B {\bf 46}, 15233
(1992). 

\bibitem{Tarucha95Yacoby96} S.~Tarucha, T.~Honda, and T.~Saku, Solid 
State Commun. {\bf 94}, 413 (1995); A.~Yacoby {\em et al.}, Phys. Rev. 
Lett. {\bf 77}, 4612 (1996).

\bibitem{Buettiker93} M. B\"{u}ttiker, J. Phys. Condensed Matter {\bf
5}, 9361 (1993). 

\bibitem{Pieper94} J.~B.~Pieper and J.~C.~Price, Phys. Rev. Lett. {\bf 
72}, 3586 (1994).

\bibitem{Schoelkopf97} R.~J.~Schoelkopf {\em et al.}, 
Phys. Rev. Lett. {\bf 78}, 3370 (1997).

\bibitem{Maslov95Ponomarenko95Safi95} D.~L.~Maslov and M.~Stone, Phys.
Rev. B {\bf 52}, R5539 (1995); V.~V.~Ponomarenko, { \em ibid} {\bf
52}, R8666 (1995); I.~Safi and H.~J.~Schulz, {\em ibid} {\bf 52},
R17040 (1995). 

\bibitem{Egger96} R.~Egger and H.~Grabert, Phys. Rev. Lett. {\bf 77},
538 (1996).

\bibitem{Nozieres} D. Pines and P. Nozi\`eres, {\it The Theory of 
Quantum Liquids} (W. A. Benjamin, N.~Y., 1966). 
 
\bibitem{Wen} Models for fractional edge states fix $g$ according 
to the relevant ground state, see X. G. Wen,
Phys. Rev. B {\bf 41}, 12838 (1990). The interaction
due to long range Coulomb forces has to be included via an 
additional interaction parameter. 

\bibitem{Alekseev97} A.~Yu.~Alekseev, V.~V.~Cheianov, and
J.~Fr\"ohlich, cond-mat/9706061 (unpublished). 

\bibitem{Cuniberti96} G.~Cuniberti, M.~Sassetti, 
and B.~Kramer, J. Phys.: Condens. Matter {\bf 8}, L21 (1996);
cond-mat/9710053 (unpublished). 

\bibitem{Ponomarenko96} V.~V.~Ponomarenko, Phys. Rev. B {\bf 54},
10328 (1995). 

\bibitem{Sablikov97} V.~A.~Sablikov and B.~S.~Shchamkhalova, Pis'ma
Zh. \'Eksp. Teor. Fiz. {\bf 66}, 40 (1997) [JETP Lett. {\bf 66}, 41
(1997)].  

\bibitem{self-cons}  The necessity of a self-consistent calculation of
the electric field was stressed for the dc problem by A.~Kawabata,
J. Phys. Soc. Jap. {\bf 65}, 30 (1996); cond-mat/9701171
(unpublished); Y.~Oreg and A.~M.~Finkel'stein, Phys. Rev. B {\bf 54},
R14265 (1996).   

\bibitem{Blanter97} Ya.~M.~Blanter and M.~B\"uttiker, cond-mat/9706070
(unpublished).

\bibitem{capacitance} 
The equivalence between the LL model with short-range interaction and
a gated wire was noted by L.~I.~Glazman, I.~M.~Ruzin, and
B.~I.~Shklovskii, Phys. Rev. B {\bf 45}, 8454 (1992); K.~A.~Matveev
and L.~I.~Glazman, Physica B {\bf 189}, 266 (1993); see also R.~Egger
and H.~Grabert, cond-mat/9709047 (unpublished).  

\bibitem{Chen} W. Chen, T. P. Smith III, M. B\"{u}ttiker, and
M. Shayegan, Phys. Rev. Lett. {\bf 73}, 146 (1994).

\bibitem{BTP93} M. B\"{u}ttiker, H. Thomas, and A. Pr\^etre,
Phys. Lett. A {\bf 180}, 364 (1993).

\end{references}
\end{document}